\documentstyle[12pt]{article}
\newread\epsffilein    
\newif\ifepsffileok    
\newif\ifepsfbbfound   
\newif\ifepsfverbose   
\newdimen\epsfxsize    
\newdimen\epsfysize    
\newdimen\epsftsize    
\newdimen\epsfrsize    
\newdimen\epsftmp      
\newdimen\pspoints     
\def\epsfbox#1{\global\def\epsfllx{72}\global\def\epsflly{72}%
   \global\def\epsfurx{540}\global\def\epsfury{720}%
   \def\lbracket{[}\def\testit{#1}\ifx\testit\lbracket
   \let\next=\epsfgetlitbb\else\let\next=\epsfnormal\fi\next{#1}}%
\def\epsfgetlitbb#1#2 #3 #4 #5]#6{\epsfgrab #2 #3 #4 #5 .\\%
   \epsfsetgraph{#6}}%
\def\epsfnormal#1{\epsfgetbb{#1}\epsfsetgraph{#1}}%
\def\epsfgetbb#1{%
\openin\epsffilein=#1
\ifeof\epsffilein\errmessage{I couldn't open #1, will ignore it}\else
   {\epsffileoktrue \chardef\other=12
    \def\do##1{\catcode`##1=\other}\dospecials \catcode`\ =10
    \loop
       \read\epsffilein to \epsffileline
       \ifeof\epsffilein\epsffileokfalse\else
          \expandafter\epsfaux\epsffileline:. \\%
       \fi
   \ifepsffileok\repeat
   \ifepsfbbfound\else
    \ifepsfverbose\message{No bounding box comment in #1; using defaults}\fi\fi
   }\closein\epsffilein\fi}%
\def\epsfclipstring{}
\def\epsfsetgraph#1{%
   \epsfrsize=\epsfury\pspoints
   \advance\epsfrsize by-\epsflly\pspoints
   \epsftsize=\epsfurx\pspoints
   \advance\epsftsize by-\epsfllx\pspoints
   \epsfxsize\epsfsize\epsftsize\epsfrsize
   \ifnum\epsfxsize=0 \ifnum\epsfysize=0
      \epsfxsize=\epsftsize \epsfysize=\epsfrsize
      \epsfrsize=0pt
     \else\epsftmp=\epsftsize \divide\epsftmp\epsfrsize
       \epsfxsize=\epsfysize \multiply\epsfxsize\epsftmp
       \multiply\epsftmp\epsfrsize \advance\epsftsize-\epsftmp
       \epsftmp=\epsfysize
       \loop \advance\epsftsize\epsftsize \divide\epsftmp 2
       \ifnum\epsftmp>0
          \ifnum\epsftsize<\epsfrsize\else
             \advance\epsftsize-\epsfrsize \advance\epsfxsize\epsftmp \fi
       \repeat
       \epsfrsize=0pt
     \fi
   \else \ifnum\epsfysize=0
     \epsftmp=\epsfrsize \divide\epsftmp\epsftsize
     \epsfysize=\epsfxsize \multiply\epsfysize\epsftmp
     \multiply\epsftmp\epsftsize \advance\epsfrsize-\epsftmp
     \epsftmp=\epsfxsize
     \loop \advance\epsfrsize\epsfrsize \divide\epsftmp 2
     \ifnum\epsftmp>0
        \ifnum\epsfrsize<\epsftsize\else
           \advance\epsfrsize-\epsftsize \advance\epsfysize\epsftmp \fi
     \repeat
     \epsfrsize=0pt
    \else
     \epsfrsize=\epsfysize
    \fi
   \fi
   \ifepsfverbose\message{#1: width=\the\epsfxsize, height=\the\epsfysize}\fi
   \epsftmp=10\epsfxsize \divide\epsftmp\pspoints
   \vbox to\epsfysize{\vfil\hbox to\epsfxsize{%
      \ifnum\epsfrsize=0\relax
        \includegraphics{#1}%
      \else
        \epsfrsize=10\epsfysize \divide\epsfrsize\pspoints
        \includegraphics{#1}%
      \fi
      \hfil}}%
\global\epsfxsize=0pt\global\epsfysize=0pt}%
{\catcode`\%=12 \global\let\epsfpercent=
\long\def\epsfaux#1#2:#3\\{\ifx#1\epsfpercent
   \def\testit{#2}\ifx\testit\epsfbblit
      \epsfgrab #3 . . . \\%
      \epsffileokfalse
      \global\epsfbbfoundtrue
   \fi\else\ifx#1\par\else\epsffileokfalse\fi\fi}%
\def\epsfempty{}%
\def\epsfgrab #1 #2 #3 #4 #5\\{%
\global\def\epsfllx{#1}\ifx\epsfllx\epsfempty
      \epsfgrab #2 #3 #4 #5 .\\\else
   \global\def\epsflly{#2}%
   \global\def\epsfurx{#3}\global\def\epsfury{#4}\fi}%
\def\epsfsize#1#2{\epsfxsize}
\textwidth6.5in
\textheight 9in\topmargin-0.7in\oddsidemargin-.0in
\baselineskip=15pt
\begin{document}
\newcommand{\bibbf}{\ninebf}
\renewenvironment{thebibliography}[1]  {
   \begin{list}{\arabic{enumi}.}     {\usecounter{enumi}
\setlength{\parsep}{0pt}      \setlength{\itemsep}{3pt}
\settowidth{\labelwidth}{#1.}      \sloppy     }}{\end{list}}

%
\newcommand{\nc}{\newcommand}
\nc{\beq}{\begin{equation}}
\nc{\eeq}{\end{equation}}
\nc{\beqa}{\begin{eqnarray}}
\nc{\eeqa}{\end{eqnarray}}
\nc{\lra}{\leftrightarrow}
\nc{\sss}{\scriptscriptstyle}
{\nc{\lsim}{\mbox{\raisebox{-.6ex}{~$\stackrel{<}{\sim}$~}}}
{\nc{\gsim}{\mbox{\raisebox{-.6ex}{~$\stackrel{>}{\sim}$~}}}
\def\lameff{\lambda_{\rm eff}}
\def\Re{{\rm Re\,}}
\def\Im{{\rm Im\,}}
\def\ns{\!\!\!\!\!\!\!\!\!\!\!\!\!\!\!\!\!\!\!\!\!\!\!\!\!\!\!\!\!\!\!\!\!\!\!}
\def\NS{\ns\ns\ns\ns}
\def\dsl{\partial\!\!\!/}
\def\Pl{P_{\sss L}}\def\Pr{P_{\sss R}}
\def\VEV#1{\langle #1 \rangle}
\def\DW{\Delta_{\rm wall}}
\nc{\rd}{{\rm d}}
\nc{\some}{\omega}
\def\section#1{{\bf #1.\ }}
\rightline{McGill/95-34}
\rightline{hep-ph/yymmddd}
\rightline{June, 1995}
\vskip .2in
\begin{center}
\renewcommand{\thefootnote}{*}
{\large{\bf Creating Matter at the Electroweak Phase
Transition}}\footnote{work done in collaboration with
Kimmo Kainulainen and Axel Vischer, presented at Great Lakes Cosmology
Workshop, Case Western Reserve
Univeristy, and MRST `95, University of Rochester, May 6-9, 1995}
\end{center}
\vskip .2truecm
\begin{center}
James M.~Cline\\
{\it McGill University, Montr\'eal, PQ H3A 2T8, Canada}
\end{center}
\vskip 0.1in
\vskip 0.5truecm
I summarize recent results in which we make quantitative predictions for the
baryon asymmetry of the universe in the charge transport mechanism of
electroweak baryogenesis.  Making favorable assumptions about the unknown
quantities relevant to the problem, we find that the mechanism is marginally
capable of creating a baryon asymmetry as large as that observed.
\vskip 0.5truecm

Although feasible explanations for the observed preponderance of matter over
antimatter have long existed, it is only since the idea of creating it
at the electroweak phase transition (EWPT) that we have the hope of testing
such theories in the laboratory.  Since the EWPT occurred when the
universe was at a temperature near 100 GeV, any new physics it might need for
producing the baryon asymmetry should be within the reach of the next
generation
of particle accelerators.

In fact one of the attractive features of electroweak baryogenesis is
that a minimal number of new physics ingredients is needed to produce
the baryon asymmetry.  Already within the standard model one has baryon
number violation via electroweak sphaleron interactions at high
temperature.  These interactions involve 9 quarks and 3  leptons, thus
violating left-handed $B+L$ by 3 units, and their rate is large up
until the EWPT.  Furthermore one has departure from thermal equilibrium
because the EWPT is  first order, proceeding by the nucleation of
bubbles of the true vacuum phase (where the Higgs field gets a VEV)
from the symmetric phase.  Thus two of Sakharov's conditions for
baryogenesis are fulfilled.  The third, CP violation, is also present
in the standard model, but most consider the form it takes there to be
too feeble for the purposes of baryogenesis, being suppressed by
factors of all the quark masses and mixing angles, and thermal damping
effects.  This shortcoming can be remedied by adding a second Higgs
doublet, for example.

A conceptually attractive way of putting these ingredients together was
proposed by Cohen, Kaplan and Nelson in 1991 \cite{CKN}, and elaborated
more recently by Joyce, Prokopec and Turok \cite{JPT}.  The idea is
that fermions from the symmetric phase have some probability of
bouncing off the expanding walls of the bubbles of true vacuum phase.
Due to CP violation in the walls, the reflection probability is
different for right-handed and left-handed fermions, and so an excess
of chirality  initially builds up in front of the moving wall. Since
sphalerons ``see'' only left-handed fermions, they  act so as
to redistribute the chirality asymmetry among all generations, reducing
the asymmetry of the original  species that was reflected most
strongly.   The sphalerons are thereby biased to change the net baryon
number away from zero.  Eventually the wall catches up to the reflected
particles which are being slowed by collisions. Once the created
baryons fall inside the bubble, they are safe from further destruction
because the sphaleron interactions are suppressed inside the bubble, if
the EWPT is strongly enough first order.

To make a quantitative estimate of the baryon production from this
mechanism we have looked at the steps in some detail, attempting to
replace assumptions made by previous investigators with  results based
on a viable model.  One focus of our work was the precise nature of CP
violation which, in the two-Higgs-doublet extension of the standard
model, comes from the relative phase $\theta(x)$ between the Higgs
fields.
Assuming that the fermion of interest couples for example to just the
second of these two fields, in order to avoid flavor-changing neutral
currents, it gets a complex, spatially varying mass term $m(z) =
\rho_2(z) e^{-i\theta(z)}$
in the vicinity of the bubble wall where the modulus $\rho_2(z)$ is
changing from zero outside the bubble to its VEV inside.  The space-dependent
 mass gives rise to quantum mechanical reflection and
transmission of the fermion, just as in a one-dimensional potential
problem, and the phase $\theta$ causes different-chirality particles to
be reflected with different probabilities.  Until now it was always
assumed that $\theta(z)$ was simply proportional to $\rho_2(z)$.  One of
our goals was to find the actual form of $\theta(z)$ in a realistic
model and to see how strongly the results depended strongly on this assumption.

More generally, we wanted to eliminate as many assumptions as possible,
such as the bubble wall width, and so it behooved us (1) to construct a
two-Higgs doublet model suitable for baryogenesis.  Subsequently we had
to (2) to find the finite-temperature form of the model; (3) solve the
equations of motion for the Higgs fields $\rho_i(z)$, $\theta(z)$ near
the bubble wall; (4) solve the Dirac equation in the background of the
Higgs fields to find the reflection probabilities; (5) determine how
the reflected fermions diffuse back into the plasma in front of the
wall; and (6) to integrate these results for the baryon asymmetry.  I
will give only the highlights of these steps here, as the details can
be found in references \cite{CKV}, \cite{CK}.

\section{The Model} To simplify the analysis we imposed the symmetry
$\phi_1\leftrightarrow\phi_2$ on the Higgs field potential, softly broken
by a dimension two operator,
\begin{eqnarray}
      V(\Phi_1,\Phi_2) &=&
      - \mu_1^2 \Phi_1^\dagger \Phi_1
       - \mu_2^2 \Phi_2^\dagger \Phi_2
      +\tilde\kappa \Phi_1^\dagger \Phi_2 + \tilde\kappa^*
      \Phi_2^\dagger \Phi_1
       + \frac{\lambda_1}{2} (\Phi_1^\dagger \Phi_1)^2
       + \frac{\lambda_2}{2} (\Phi_2^\dagger \Phi_2)^2
      \phantom{hannatytto} \nonumber \\
      & &+ h_1(\Phi_1^\dagger \Phi_2)(\Phi_2^\dagger \Phi_1)
       + h_2(\Phi_1^\dagger \Phi_1)(\Phi_2^\dagger \Phi_2)
      + h_3\left((\Phi_1^\dagger \Phi_2)^2 + (\Phi_2^\dagger \Phi_1)^2)
      \right)
\label{treepot}
\end{eqnarray}
Because of the symmetry, both Higgs fields will get VEV's whose moduli
are equal, $\rho_1 = \rho_2 \equiv \rho$, and so we have reduced the
problem to two fields, $\rho(z)$ and $\theta(z)$.   This also
circumvents the complexities of a two-stage phase transition, where one
field gets its VEV before the other.

The two terms responsible for CP violation at zero temperature are those
with the coefficients $\kappa$ and $h_3$.  If either one is missing (or if
their phases happen to be matched just right), all the couplings in the
potential can be made real by a global field redefinition.

\section{Effective Potential} We used the ring-improved
finite-temperature effective potential, which is obtained from the
usual one-loop finite-$T$ result by substituting the
temperature-corrected masses for the zero-temperature ones.  It is
important to avoid the unitary gauge in this construction, as it is
known to give unreliable results near the critical temperature of the
phase transition \cite{ABV}.

\section{CP-Violating Phase Profile}  The shape of the bubble wall is given by
a familiar form $\rho = \rho_c g(z)$, $g(z) =
\frac12(1+\tanh(z/\Delta))$, where $\rho_c$ is the Higgs field VEV
inside the bubble and $\Delta$ is the width of the wall.  One finds that
the equation for the shape of the phase $\theta(z)$ is
\beq
	\Delta^2\partial_z^2\theta + 4\Delta(1-g)\partial_z\theta +
	V(\theta) = 0,
\eeq
where $V(\theta)$ is a potential of the form
\beq
	V(\theta) = 2|\kappa|\Delta^2\sin(\theta-\delta) +
	\frac{32}{h_3}{\lambda_{\rm eff}}\, g^2 \sin2\theta,
\eeq
and we have assumed that $\theta$ is small so that the back-reaction of
$\theta$ on the equation for $\rho$ could be neglected.
The CP-violating phase is $\delta$, which is defined to be the phase of
$-\kappa$ after doing the redefinition of the Higgs fields in
(\ref{treepot}) needed to make the $h_3$ coupling real.  $\lambda_{\rm
eff}/4$ is the effective quartic coupling in the potential for $\rho$
after one has substituted the parametrization $\phi_i = \rho_i e^{\pm
i\theta/2}/\sqrt{2}$ into the potential (\ref{treepot}).   For most values
of  $\kappa$, the amount $\Delta\theta$ by which $\theta$ changes in
going between the broken and symmetric phases is proportional to
$\delta$. However it is interesting that $V(\theta)$ has nontrivial
minima even when $\delta = 0$ if $\kappa$ is sufficiently negative, due
to the changing background field $g(z)$.  Thus CP can be spontaneously
violated during the phase transition, even though there is no sign of
it in the potential at zero temperature.

Some solutions for $\theta(z)$ are shown in figure 1, which
demonstrates how they differ from the commonly used ansatz (labeled
``tanh'') for different values of $\kappa$.  The curves are distinguished by
a parameter $\zeta = -\kappa/m^2_h$, where $\kappa$ is by convention
negative and $m_h$ is the mass of the lightest Higgs boson, assumed to
be 60 GeV. The light mass helps make the sphaleron interactions as slow
as possible inside the bubble wall, so that the baryon asymmetry is not
destroyed after it is created.

\epsfbox{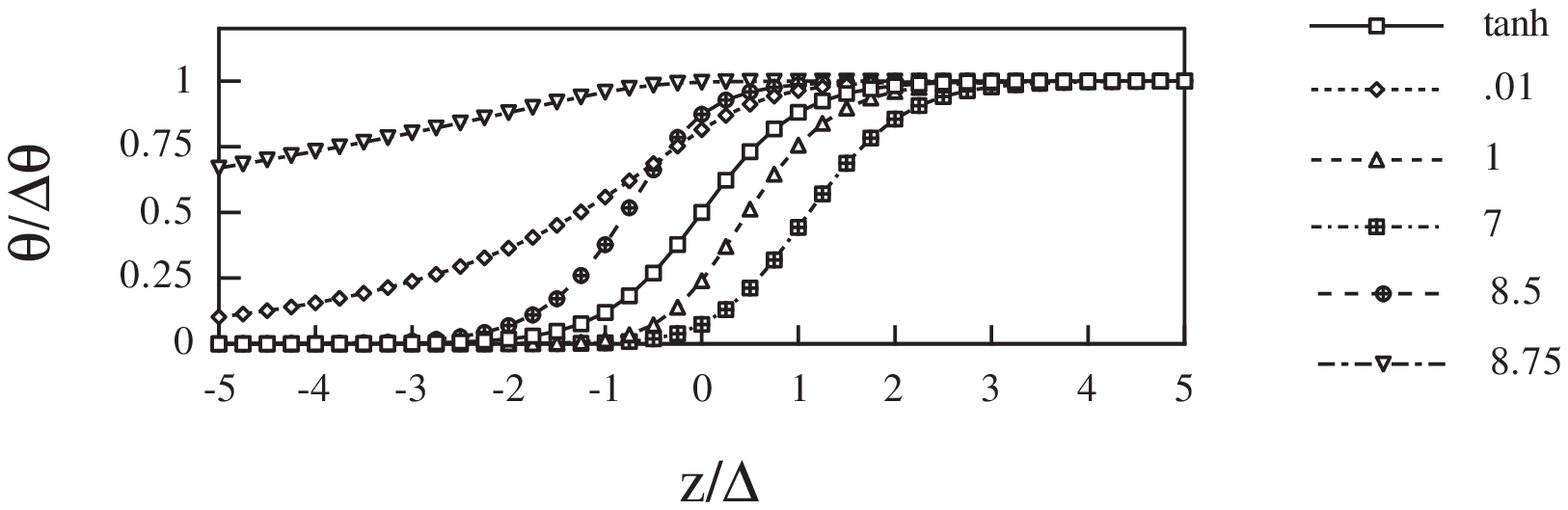}
\noindent{{\bf Figure 1.\ }
 Solutions for the relative phase of the two Higgs fields
at the bubble wall.}\vskip 0.5cm

\section{Dirac Equation}  We solved the Dirac equation using two
methods, first by directly integrating it and second using the formula
of Funakubo {\it et al.\ }\cite{Fun}, which treats $\theta$ as a small
perturbation.  We were thus able to verify the latter formula
independently.  The goal was to solve for the difference in probabilities
for right-handed fermions and antifermions to reflect from the wall
into their opposite-chirality counterparts.  The difference is called
$\Delta R\equiv R_{R\to L} - R_{\bar R\to\bar L}$.
A convenient and fairly accurate parametrization of $\Delta R$ as a
function of the fermion momentum is
\beq
\Delta R(p_z)  = 
		 |\Delta R|_{\rm max} e^{-p_z/w}, \qquad p_z > m. 
\label{fit}
\eeq
For $p_z < m$,  $\Delta R$ vanishes because such particles do not have
enough energy to penetrate into the broken phase, since they go from
being massless to having a mass, and so both particles and
antiparticles are completely reflected.  The height and width of the
exponential depends on the mass and the width of the bubble wall
through their product, $\xi = m\Delta$.

In figure 2 the comparison between the actual functions $\Delta R(p_z)$
and the fit (\ref{fit}) is shown.  Although the fit is not perfect, it
is only necessary to match the area under the curves because they will
ultimately be integrated over $p_z$, combined with other functions that
are rather flat on the scale over which $\Delta R(p_z)$ is changing.

\epsfbox{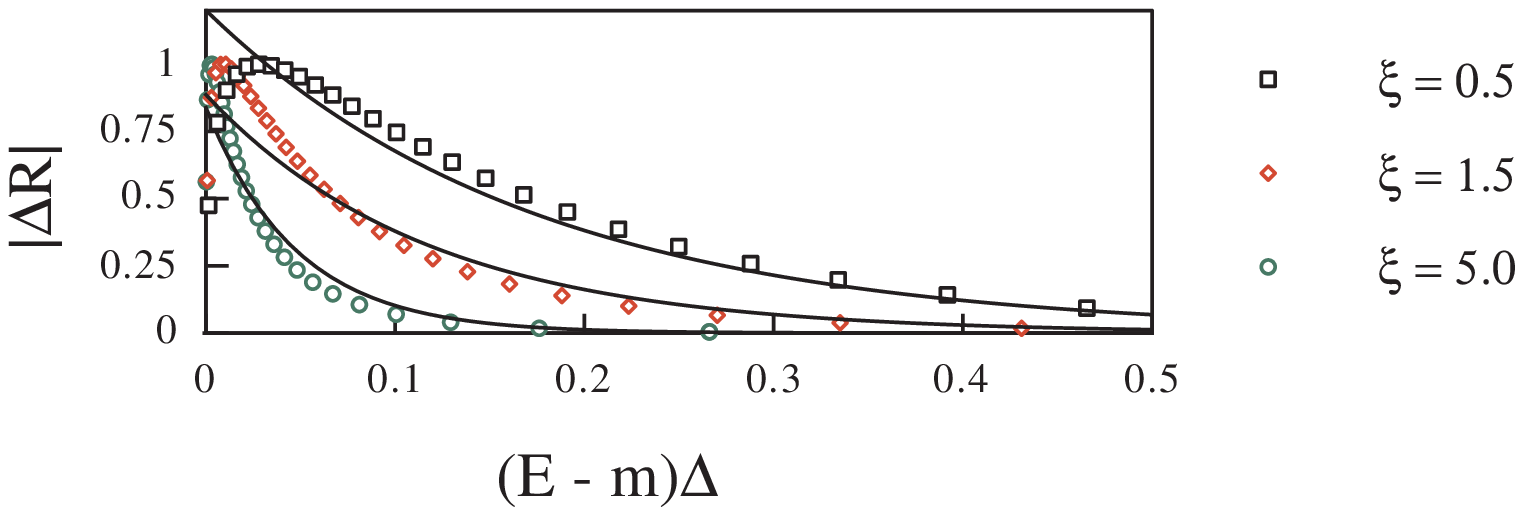}
\noindent{{\bf Figure 2.}\ Reflection probability asymmetry for different
fermion masses ($\xi = m\Delta$), together with the fits (\ref{fit}).}
\vskip 0.5cm

\epsfbox{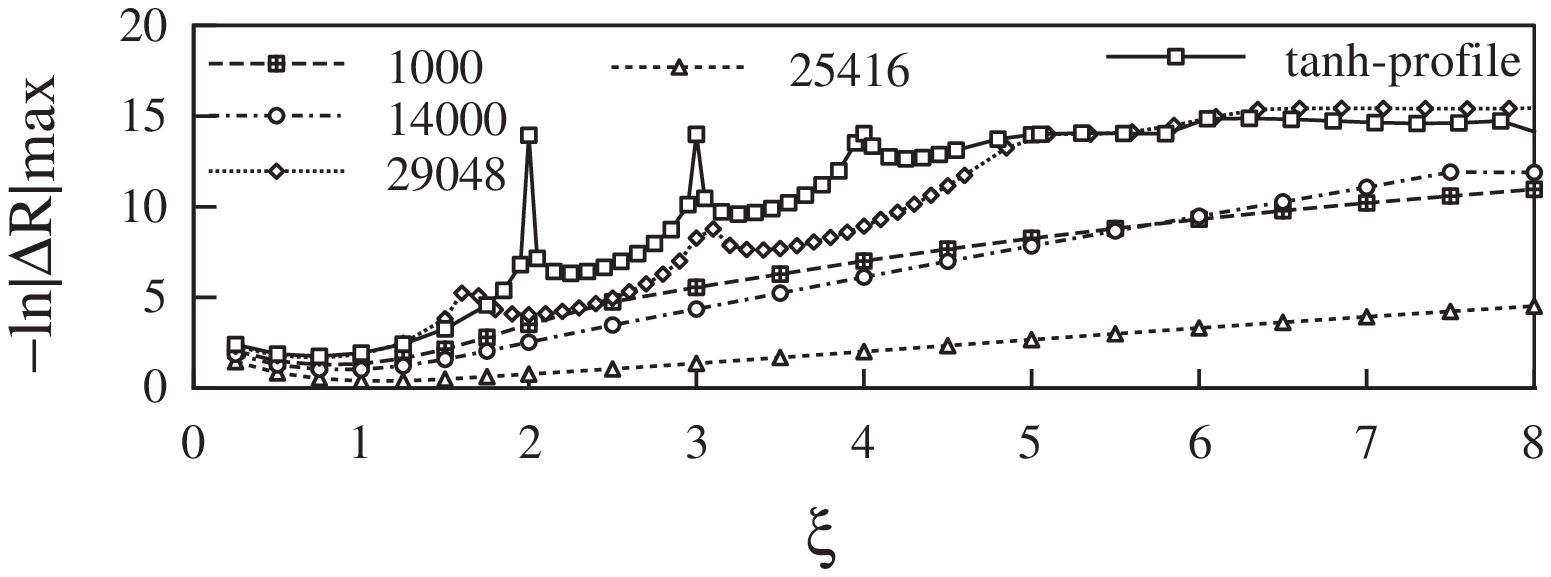}
\noindent{{\bf Figure 3.}\
Amplitude of the reflection asymmetry as a function of
fermion mass.  Curves are labeled by the value of $-\kappa$, chosen to
coincide with some of the values used in fig.\ 1}\vskip 0.5cm

In figure 3 we show how the amplitude of $\Delta R$ depends on the
fermion mass parameter $\xi= m\Delta$.  The square boxes indicate the
case of the ansatz $\theta(z) \propto \rho(z)$ (called the ``tanh
ansatz'').  The spikes are where $|\Delta R|_{\rm max}$ goes through
zero, which occurs near integer values of $\xi$.  ($\Delta R(p_z)$ does
not actually vanish identically at these values of $\xi$; rather the
parametrization (\ref{fit}) breaks down, and the area under the curve
$\Delta R(p_z)$ vanishes; nevertheless its maximum value is still
highly suppressed for integer values of $\xi$.)   The other curves show
the case of the real solutions for various values of $-\kappa$, the
Higgs potential parameter.  Typically these do not display the spiky
behavior of the tanh ansatz, unless the corresponding $\theta$ profiles
are very close in shape to that of the tanh ansatz.  One notices that
the variations between the actual solutions and the ansatz are most
pronounced for $\xi > 1$.  It so happens that the actual spectrum of
fermions in the standard model is $\xi = 0.12$, $0.33$ and $11.7$ for
the tau lepton, bottom quark and top quark, respectively.  As will be
explained below, the top quark is irrelevant for baryogenesis in this
model, so for the relevant fermions, the difference between the ansatz
and the actual solutions is small.

We have also examined how the width $w(\xi)$ of the $\Delta R$ profiles
varies as a function of the fermion mass.  This width is defined to be
the area under the curve $\Delta R(p_z)$ divided by the maximum value
of $\Delta R$ discussed above.  The dependence is shown for typical
values of the model parameters in figure 4.  In contrast to the
amplitude $\Delta R_{\rm max}$, we find that $w(\xi)$ is virtually
independent of the potential parameter $\kappa$.

\epsfbox{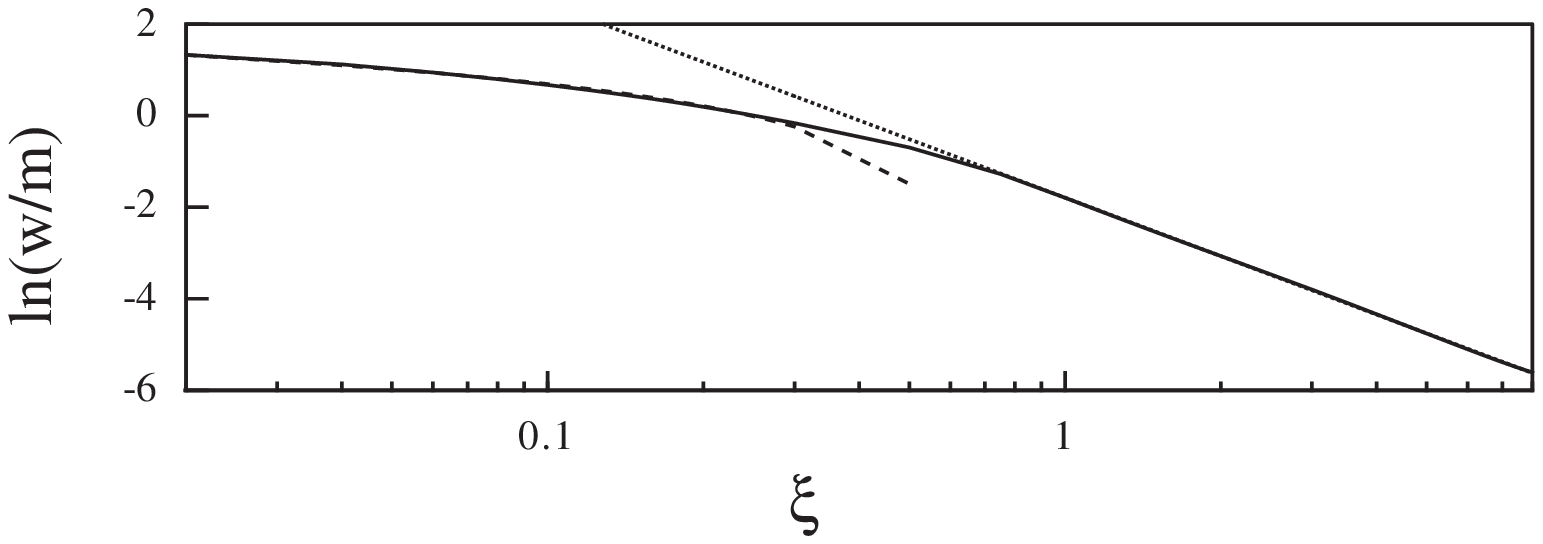}
\noindent{{\bf Figure 4.}\  Momentum-space width of the reflection asymmetry
as a function of fermion mass.}\vskip 0.5cm

An interesting but (as it turns out) inessential complication in solving the
Dirac equation is that for very low momenta, the dispersion relations of the
fermion are altered from their zero-temperature form, $E=p$ in the symmetric
phase or $E= (p^2+m^2)^{1/2}$ in the broken phase.  The left- and right-handed
particles get thermal contributions to their effective masses $\omega_L$ and
$\omega_R$, and the dispersion relations are altered to \cite{FS}
\beqa
	E = \omega_{L(R)} \pm p/3, && \hbox{\ symmetric phase;}\\
	E = \frac{\omega_L+\omega_R}{2}\pm\sqrt{\left(\frac{
	\omega_L-\omega_R}{2}\pm\frac{p}{3}\right)^2 +\frac{m^2}{4}},
	&& \hbox{\ broken phase},
\eeqa
in the regions of momentum space where $|p| < \omega_i$.  The Dirac equation
must be accordingly altered.  However it was found that the small-momentum
region makes a subdominant contribution to the baryon asymmetry compared to the
region where the usual dipsersion relations prevail.

\section{Equilibration and Transport} After being reflected the
fermions diffuse into the symmetric phase, but before going far,
interactions change their chemical composition.  The most important
of these are the strong sphalerons \cite{GetS}, the QCD analog of electroweak
sphalerons. They cause chirality-changing transitions among the quarks
and because they are fast, they give rise to the equilibrium condition
that left-handed and right-handed baryon number are equal, $B_L =
B_R$.  On the other hand the initial condition at the wall was that the
fermions had no net baryon number, $B_L + B_R = 0$, because it was
assumed to be zero at the outset and we find that the weak sphalerons
are too slow to have yet changed the net baryon number.  The solution
to these equations is obviously the trivial one, $B_L = B_R = 0$.  But
there are thermal corrections which make the actual solution not quite
zero; instead what happens is that the asymmetries $B_L$ and $B_R$ get
reduced by a factor $\sim 20$ from their initial values at the wall.
Of course only quarks undergo this suppression because leptons have no
strong interactions.  The other significant interaction is the Higgs
coupling to top quarks, which we took into account in the above
estimates.  For the lighter fermions the Higgs coupling is too small to
be important on the time-scale for fermions to diffuse in front of the
wall.

The diffusion process itself has been modeled using the diffusion equation
for the particle densities,
\beq
	{\partial n\over\partial t} - D{\partial^2n\over\partial x^2}
	= 0,
\eeq
and an approximation to the Boltzmann equation for the distribution functions,
\beq
	{\partial f\over \partial t} + \vec v \cdot {\partial f\over \partial
	\vec x} = \widetilde D {\partial\over\partial\vec p}\cdot\left(
	{\partial f\over\partial\vec p} + \beta\vec v f\right),
\eeq
known as the Fokker-Planck equation \cite{jc}.
It was argued that the latter more
correctly describes the present situation because of the strong momentum
dependence of the distribution functions, owing to the smallness of the
momentum space width of the reflection probabilities discussed above,
$w\ll T$.  It can be shown in the one-dimensional case that the solution to the
Fokker-Planck equation is only compatible with that of the diffusion equation
if $f(p)$ evaluated at the wall falls off with $p$ like $e^{-\beta p}$, namely
a thermal distribution.  But one must remember that here $f$ represents the
asymmetry between left-handed particles and antiparticles due to the
CP-violating reflections, and this asymmetry has a distribution which is far
from thermal.  In fact it should go like $\Delta R(p)\sim e^{-p/w}$, and hence
the importance of the fact that $w\ll T$.  The Fokker-Planck equation gives
qualitatively different results; it predicts that the integrated asymmetry in
front of the wall should be independent of the wall velocity (for a range of
velocities $v > 0.01$) and be cubic in the larger of $w$ or $m$ \cite{jc}.

Even though these parametric dependences are known, an inconsistency
was recently found in the numerical evaluation of integrals in
ref.~\cite{jc}, which are needed to find the overall normalization of
the integrated asymmetry in front of the wall in three dimensions.
Pending the resolution of this problem, I will fall back upon the
simpler diffusion equation and defer discussion of the differences
between the two approaches to a future publication.  Then the solution
for the chiral asymmetry in front of the wall is a simple exponential,
$n(z) = v^{-1}J(0)e^{-vz/D}$, where the initial flux $J(0)$ at the wall
can be computed from the reflection asymmetry $\Delta R$ and the
distribution functions for incident and reflected particles.

\section{The Baryon Asymmetry} To compute the baryon asymmetry at a
given position $z$ in space, one must integrate the rate of baryon
number violation by sphalerons from very early times (we are assuming a
steady-state distribution of chirality in front of the wall) until the
time the wall passes by that position.  The local rate of baryon violation is
proportional to the chiral asymmetry, and its time integral can be converted to
the spatial integral of the asymmetry in front of the wall.  The result
for the ratio of baryon number to entropy of the universe is
\beqa
        {n_B\over s} &=&  3\times 10^{-3}\alpha_W^4\; D \;{m^2\over T_c^2}
         \; \Delta\theta \; A(\xi)\; w(\xi) \;
          r(w/m) \; /\; v^2; \\
\label{result}
	r(w/m) &=&   1 + 2w/m + 2w^2/m^2+ K_2(m/w)e^{m/w}.\nonumber
\eeqa
not counting the suppression factor of $\sim 20$ for contributions from the
reflected quarks. Because of this factor and since the diffusion coefficient
$D$ is much smaller for quarks than for leptons, one finds that the tau lepton
makes the dominant contribution \cite{JPT}, resulting in ${n_B/s} \simeq 1
\times 10^{-12} \; \Delta\theta /v^2$, which is compatible with the number
determined from nucleosynthesis, $4-6\times 10^{-11}$, but only if the wall
velocity is near $0.1$, the lower end of its expected range, and if CP is
maximally violated.  This assumes the rate per unit volume of sphaleron
interactions is $(\alpha_W T)^4$, but it might be larger by a factor of 10,
making a wall velocity of $v = 0.3$ acceptable.

Should the CP-violating phase be too small however, there may still be
ways of increasing the result; for example ref.~\cite{JPT} suggests
that the VEV of the Higgs field coupling to the tau lepton may be
larger relative to the r.m.s.\ VEV during the phase transition than at
$T=0$, so that the effective fermion mass is larger than one would
infer from the zero-temperature fermion mass spectrum.  Whether this
can be made to happen in an actual model has not yet been demonstrated,
and it would certainly involve a two-stage phase transition with its
potential attendant complications.  Another possibility is to alter the
model so as to increase the ratio of the VEV to the temperature during
the phase transition, since this would increase the ratio $m/T$ in
(\ref{result}).  This would also have the advantage of slowing the
sphaleron interactions in the phase inside the bubbles, as one wants
for preserving the baryon asymmetry.

 I would like to thank Axel Vischer for his
generous assistance in preparing the figures.  This work was partially
supported by NSERC of Canada and FCAR du Qu\'ebec.  \vskip 0.5cm

 {\large\bf References}

\end{document}